\documentclass[11pt]{article}
\usepackage{amsmath}
\begin{document}
\topmargin -1.4cm
\oddsidemargin -0.8cm
\evensidemargin -0.8cm
\def\g#1{{\scriptstyle (\! #1 \! )}}
\def\Sg#1{{\scriptstyle [\! #1 \! ]}}
\def\Ng#1{{\scriptstyle ~\! #1 \! ~}}
\def\gg#1{{\scriptscriptstyle (\! #1 \! )}}
\def\Sgg#1{{\scriptstyle [\! {\scriptscriptstyle #1} \! ]}}
\def\Ngg#1{{\scriptstyle ~\! {\scriptscriptstyle #1} \! ~}}
\def\IUg[#1]{ {\mathsf{I}}^{\scriptstyle [\! #1 \! ]}{}}
\def\IDg[#1]{ {\mathsf{I}}_{\scriptstyle [\! #1 \! ]}{}}
\def\IUgg[#1]{{\mathsf{I}}^{\scriptscriptstyle [\! #1 \! ]}{}}
\def\IDgg[#1]{{\mathsf{I}}_{\scriptscriptstyle [\! #1 \! ]}{}}
\def\xk{{\mathbf{x},k}}
\def\xp[#1]{{{\mathbf{x}}\!+\!#1}}
\def\xm[#1]{{{\mathbf{x}}\!-\!#1}}
\def\yp[#1]{{{\mathbf{y}}\!+\!#1}}
\def\ym[#1]{{{\mathbf{y}}\!-\!#1}}
\def\DEFxmk{{{\mathbf{x}}-a{\mathbf{e}}_k,k}}
\def\DEFxpk{{{\mathbf{x}}+a{\mathbf{e}}_k,k}}
\def\Rgen[#1]#2#3{{ 
        D^{ {\scriptstyle (\! #1 \! )}#2}_{~~~#3} }}
\def\gUP[#1]#2{{   g_{{[\! #1 \! ]}}^{{#2 }} }}
\def\gDOWN[#1]#2{{ g^{{[\! #1 \! ]}}_{{#2 }} }}
\def\ThreeJUP[#1]#2{  \left|{}_{[\!#1\!]}^{#2} \right| }
\def\ThreeJDOWN[#1]#2{\left|{}^{[\!#1\!]}_{#2} \right|}
\def\ThreeJ#1#2#3#4#5#6{{\left(\!
\begin{array}{ccc} #1 & #2 & #3 \\
                   #4 & #5 & #6
\end{array} \! 
\right)}}
\def\SixJ#1#2#3#4#5#6{{\left\{\!
\begin{array}{ccc} #1 & #2 & #3 \\
                   #4 & #5 & #6
\end{array} \! 
\right\}}}
\def\DELTA#1{{
     \!\begin{array}{c}\setlength{\unitlength}{.5 pt}
     \begin{picture}(35,25)
        \put(15, 0){\line(0,1){10}} \put(20, 0){\line(0,1){10}}
        \put(15, 0){\line(1,0){ 5}} \put(15,10){\line(1,0){ 5}}
        \put(15,25){\line(1,0){5}}  \put(15,15){$\scriptstyle {#1}$}
        \put(15,15){\oval(20,20)[l]}\put(20,15){\oval(30,20)[r]}
     \end{picture}\end{array} \!
}}
\def\THETA#1#2#3{{
     \begin{array}{c}\setlength{\unitlength}{.5 pt}
     \begin{picture}(40,40)
        \put(18,32){$\scriptstyle {#1}$}
        \put( 0,15){\line(1,0){40}} \put(18,17){$\scriptstyle {#2}$}
        \put(20,15){\oval(40,30)}   \put(18, 2){$\scriptstyle {#3}$}
        \put( 0,15){\circle*{3}}    \put(40,15){\circle*{3}}
     \end{picture}\end{array}
}}
\def\TET#1#2#3#4#5#6{{
     \begin{array}{c}\setlength{\unitlength}{.8 pt}
        \begin{picture}(50,30)
        \put( 0,15){\line(1,-1){15}} \put(0,22){${\scriptstyle {#2}}$}
        \put( 0,15){\line(1, 1){15}} \put(0, 0){${\scriptstyle {#1}}$}
        \put( 0,15){\circle*{3}}
        \put(30,15){\line(-1, 1){15}} \put(28,20){${\scriptstyle {#4}}$}
        \put(30,15){\line(-1,-1){15}} \put(28, 2){${\scriptstyle {#5}}$}
        \put(30,15){\circle*{3}}
        \put( 0,15){\line(1,0){30}} \put(12,16){${\scriptstyle {#6}}$}
        \put(15,30){\line(1,0){25}} \put(15,30){\circle*{3}}
        \put(15, 0){\line(1,0){25}} \put(15, 0){\circle*{3}}
        \put(40, 0){\line(0,1){30}} \put(42,12){${\scriptstyle {#3}}$}
    \end{picture}\end{array}
}}
\def\NCA{\em Nuovo Cimento}
\def\NIM{\em Nucl. Instrum. Methods}
\def\NIMA{{\em Nucl. Instrum. Methods} A}
\def\NPB{{\em Nucl. Phys.} B}
\def\PLB{{\em Phys. Lett.}  B}
\def\PRL{\em Phys. Rev. Lett.}
\def\PRD{{\em Phys. Rev.} D}
\def\ZPC{{\em Z. Phys.} C}
\def\Journal#1#2#3#4{{#1} {\bf #2}, #3 (#4)}

\title{Matrix elements of the plaquette operator of Lattice Gauge Theory }

\author{
Giuseppe Burgio, Roberto De Pietri \\[2mm]
Dipartimento di Fisica, Universit\`a degli Studi di Parma and \\
I.N.F.N.\ gruppo collegato di Parma, \\
Parco Area delle Scienze 7/a, I-43100 Parma, Italy \\
E-mail: burgio@pr.infn.it, depietri@pr.infn.it\
\\[4mm] 
H. A. Morales-T\'ecotl \\[2mm]
Departamento de F\'\i sica, Universidad Aut\'onoma
Metropolitana Iztapalapa, \\ 
A. Postal 55-534, 09340  M\'exico, D.F.
\\[4mm] 
L. F. Urrutia, and  J. D. Vergara \\[2mm]
Departamento de F\'\i sica de Altas Energ\'\i as, Instituto
de Ciencias Nucleares,  \\
Universidad Nacional Aut\'onoma de M\'exico, 
A. Postal 70-543, 04510 M\'exico D.F.
}


\maketitle

\begin{flushleft}
University of Parma PREPRINT UPRF-99-19, November 1999. \\
Talk presented at: XVII Autumn school "QCD: perturbative or non-perturbative?" 
Istituto Superior Tecnico, Lisboa, 29/9-4/10/1999.
\end{flushleft}

\begin{abstract}
We show that in the {\it spin-network} basis it is possible to compute
the matrix elements of any given operator of the Hamiltonian formulation of 
Lattice Gauge Theory (LGT). We give the explicit calculation 
for the case of the plaquette operator. 
\end{abstract}


\section{Introduction} 
\protect\label{sec:INTRO}

In recent papers\cite{BurgioEtAl:1999,BurgioEtAl:1999bis} we proposed
a group theoretical description of the Hilbert space for lattice gauge theories (LGT)
in the
Hamiltonian framework\cite{Kogut:1975}.  
This approach, based on representation theory, allows to overcome the
problem of selecting the gauge invariant Hilbert space. In particular, 
the difficulty of explicitly solving Mandelstam's identities\cite{Mad}
in the context of Wilson loops\cite{Wilson:1974} is 
circumvented. Moreover, such an approach yields a general setting for the
computation of the matrix elements of the relevant operators. 

We will briefly sketch the main concepts underlying our construction and 
present a physically interesting application as an example.   

\section{The spin network basis}

In $(d\!+\!1)$ dimensions the configuration space of LGT is defined
associating gauge field variables $U_k({\mathbf{x}})\in G$ to each
link $({\mathbf{x}},{\mathbf{x}}+a{\mathbf{e}}_k)$ of a hypercubic
periodic lattice of period $a L$, with $L$ a positive integer.

The corresponding quantum Hilbert space ${\mathcal{H}}$ is given by
the gauge invariant square integrable functions
$\psi(U)=\psi(U^\gamma)=\psi(\{ U_k^\gamma(\mathbf{x}) \})$ on the
tensor product of $d \cdot L^d$ copies of the gauge group $G$. Gauge
transformations act as $U_k({\mathbf{x}}) \longrightarrow
U_k^\gamma({\mathbf{x}}) = \gamma^{-1}({\mathbf{x}}+a{\mathbf{e}}_k)
U_k({\mathbf{x}}) \gamma({\mathbf{x}})$.  The variables conjugated to
$U_k(\mathbf{x})$ are the outgoing/ingoing electric fields
$E^\alpha_{\pm\! k}({\mathbf{x}})$ at the lattice point
${\mathbf{x}}$ in the directions $ {\mathbf{e}}_{\pm\!k}$.

A classical result of representation theory\cite{BookGroup} gives a way of
constructing a basis of such Hilbert space.  In fact, the set
${\mathcal{RG}} = \{ {\mathcal{R}}^{j} ~| j\in J[G] \}$ of all the
unitary inequivalent representations of a compact group $G$ is
numerable and all the representations ${\mathcal{R}}^{j}$ are finite
dimensional. They are defined on the Hilbert space ${\cal H}^j$.
Choosing an orthonormal basis for each representation
${\mathcal{R}}^{j}$, the matrix elements 
of a unitary operator $T(U)$ become $\Rgen[j]{\alpha}{\beta}(U)$
($\alpha,\beta=1,\ldots,\mathrm{dim}({\mathcal{H}}^j)$) of all the
representations ${\mathcal{R}}^{j}$ are a numerable orthonormal basis
of ${\mathcal{L}}^2[G,dU]$.  This result, known as the Peter-Weyl
theorem, implies that each vector of $\mathcal{H}$
can be written as
\def\SHORTCUTforC{ {c^{\g{j_\gg{1}\cdots j_\gg{N_{lk}}}
\beta_\gg{1}\cdots\beta_\gg{N_{lk}}
}_{~~~~~~~~~~~\alpha_\gg{1}\cdots\alpha_\gg{N_{lk}}} }}
\def\SHORTCUTforCbar{ {c^{\g{j_\gg{1}\cdots j_\gg{N_{lk}}}
\bar{\beta}_\gg{1}\cdots\bar{\beta}_\gg{N_{lk}}
}_{~~~~~~~~~~~\bar{\alpha}_\gg{1}\cdots\bar{\alpha}_\gg{N_{lk}}} }}
\begin{eqnarray} \label{eq1}
&& \psi(U) = \prod_{\mathbf{x}} \prod_{k=1}^{d} \sum_{j_{\mathbf{x}}^k
\in J[G]} \sum_{\alpha_{\mathbf{x}}^k,\beta_{\mathbf{x}}^k=1}^{
{{\mathrm{dim}}(j_{\mathbf{x}}^k)}} \bigg[
     ~\Rgen[j_{\mathbf{x}}^k]{\alpha_{\mathbf{x}}^k}{\beta_{\mathbf{x}}^k}(U)
     \times \SHORTCUTforC \bigg],
\end{eqnarray}
where only gauge invariant combinations should be taken into account.

The implementation of gauge invariance turns into a set of constraints
on the coefficients $c$. The $c$'s factorize as products of group
invariant tensors (intertwining operators) associated to the different
lattice sites ${\mathbf{x}}$.  By definition, an operator ${\mathsf
I}$ connecting the Hilbert space of two representations, $\mathcal{R}$
and ${\mathcal{R}}'$, is an intertwining operator if ${\mathsf{I}}
\cdot T(U) = T'(U) \cdot {\mathsf{I}}$, for every $U$ in $G$.  The set
of all intertwining operators
${\mathcal{I}}({\mathcal{R}},{\mathcal{R}}')$ is a vector subspace of
all the linear operators connecting the Hilbert space of the two
representations ${\mathcal{R}}$ and ${\mathcal{R}}'$.  This gives the
coordinate free definition of the generalized Clebsh-Gordan
coefficients of Yutsis-Levinson-Vanagas which are
the matrix elements of these operators on the chosen basis. The
integral of the product of $K$ representations decomposes according to
\begin{equation}
 \int\!\! dU \prod_{k=1}^K \Rgen[j_k]{\alpha_k}{\beta_k}(U) \!=\!
 \sum_{\pi} \frac{\IUg[\pi]{}^{(j_1\ldots j_K)}_{\beta_1\ldots\beta_K}
 \IDg[\pi]{}_{(j_1\ldots j_K)}^{\alpha_1\ldots\alpha_K}
 }{\IUg[\pi]{}^{(j_1\ldots j_K)}_{\gamma_1\ldots\gamma_K}
 \IDg[\pi]{}_{(j_1\ldots j_K)}^{\gamma_1\ldots\gamma_K} },
\label{eq2}
\end{equation}
where $\IDg[\pi]{}_{(j_1\ldots j_K)}\in
{\mathcal{I}}({\mathcal{R}}^{j_1}
\otimes\ldots\otimes{\mathcal{R}}^{j_K},\emptyset)$,
$\IUg[\pi]{}^{(j_1\ldots j_K)}$ is its adjoint and the sum is extended
over a complete orthogonal basis of ${\mathcal{I}}({\mathcal{R}}^{j_1}
\otimes\ldots\otimes{\mathcal{R}}^{j_K},\emptyset)$.
\begin{figure}[t]\protect\label{fig}
\def\BasicBlockTwoDim#1#2#3#4#5{{ \mbox{{\setlength{\unitlength}{1pt}
\begin{picture}(70,70)
   \put(35,35){\circle{25}} \put(30,30){\line( 1,1){10}}
   \thicklines 
   \put( 0,35){\line(6,-1){30}} \put(70,35){\line(-6, 1){30}} \put(35,
   0){\line(-1,6){5}} \put(35,70){\line( 1,-6){5}}
   \thinlines 
   \put(30,30){\circle*{3}} \put(40,40){\circle*{3}}
   \thinlines 
   \put( 7,38){$\scriptstyle #1$} \put(36,10){$\scriptstyle #2$}
   \put(33,29){$\scriptstyle #3$} \put(52,42){$\scriptstyle #4$}
   \put(40,53){$\scriptstyle #5$}
\end{picture}}}  }}
\centerline{
\mbox{\setlength{\unitlength}{1pt}
\begin{picture}(140,140)
   \put( 0, 0){\BasicBlockTwoDim{}{}{
   \pi_{\mathbf{x}}^1}{j_{\mathbf{x}}^1}{j_{\mathbf{x}}^2}} \put(70,
   0){\BasicBlockTwoDim{}{j^2_{\xp[1\!-\!2]}}{
   \pi^1_{\xp[1]}}{j^1_{\xp[1]}}{j^2_{\xp[1]}}}
   \put( 0,70){\BasicBlockTwoDim{j_{\xm[1\!+\!2]}^1}{}{
   \pi^1_{\xp[2]}}{j^1_{\xp[2]}}{j^2_{\xp[2]}}}
   \put(70,70){\BasicBlockTwoDim{}{}{\pi^1_{\xp[1\!+\!2]}}{}{}}
\end{picture}}
~~~ ~~~ \mbox{\setlength{\unitlength}{1pt}
\begin{picture}(140,140)
   \put( 10, 75){$\scriptstyle j_\xm[1]^1 $} \put(120,
   75){$\scriptstyle j_\mathbf{x}^1$} \put( 25, 30){$\scriptstyle
   j_\xm[2]^2 $} \put(110,105){$\scriptstyle j_\mathbf{x}^2$} \put(
   75, 10){$\scriptstyle j_\xm[3]^3 $} \put( 75,120){$\scriptstyle
   j_\mathbf{x}^3$} \put( 34, 60){$\scriptstyle \pi_\mathbf{x}^1$}
   \put( 48, 60){$\scriptstyle \pi_\mathbf{x}^2$} \put( 58,
   48){$\scriptstyle \pi_\mathbf{x}^3$}
   \thinlines 
   \put(45,55){\circle*{3}} \put(45,55){\line(0,1){15}}
   \put(45,70){\circle*{3}} \put(45,55){\line(1,0){10}}
   \put(55,55){\circle*{3}} \put(70,55){\oval(50,20)[bl]}
   \put(70,45){\circle*{3}}
   \thicklines 
   \put( 0,70){\line(1,0){140}} \put(35,35){\line(1,1){10}}
   \put(50,50){\line(1,1){15}} \put(105,105){\line(-1,-1){30}}
   \put(70, 0){\line(0,1){65}} \put( 70,140){\line( 0,-1){65}}
\end{picture}}}
\caption{Graphical representation of (a possible) decomposition of a
  vertex in a two dimensional (left) and in a three dimensional
  (right) lattice. The solid dots denote the presence of a Wigner 3J symbol.}
\label{fig:VERTEX}
\end{figure}
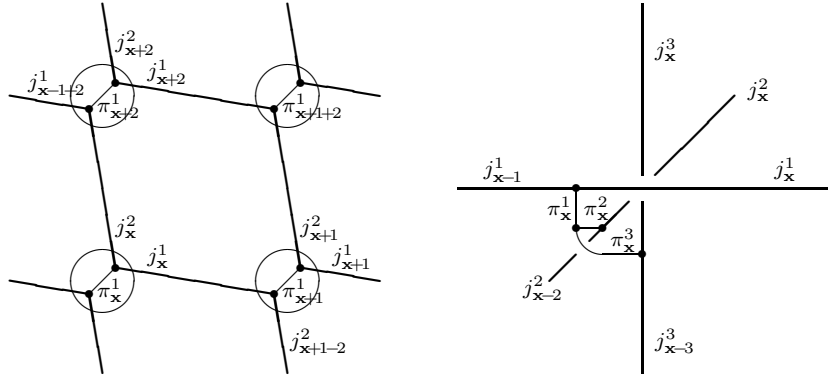

Summarizing, Peter-Weyl theorem together with gauge invariance lead to the {\it
spin network} basis elements
\begin{eqnarray} 
&& \psi_{\vec{\jmath},\mathbf{\vec{\pi}}}(U) = \prod_{{\mathbf{x}}}~
\prod_{k=1}^{d}
\sum_{\alpha_{{\mathbf{x}}}^k,\beta_{{\mathbf{x}}}^k=1}^{
{{\mathrm{dim}}(j_{{\mathbf{x}}}^k)}} \bigg[
\label{eq3} 
   ~\Rgen[j_{\mathbf{x}}^k ]{{\alpha}_{\mathbf{x}}^k}{
   {\beta}_{\mathbf{x}}^k}(U_k({\mathbf{x}})) \cdot
   \mathsf{I}_{{\mathbf{x}}}^{\Sg{{\mathbf{\pi}}_{\mathbf{x}}}}
   {}^\g{j_{\xm[1]}^1,\ldots,j_{\xm[d]}^d}_{
   \alpha_{\xm[1]}^1,\ldots,\alpha_{\xm[d]}^d}
   {}_\g{j_{\mathbf{x}}^1,\ldots,j_{\mathbf{x}}^d}^{
   \beta_{\mathbf{x}}^1,\ldots,\beta_{\mathbf{x}}^d} \bigg]~.
\end{eqnarray}
The only non trivial part in this characterization is the choice of a
convenient basis for the intertwining matrices. A natural choice is to
use an orthonormal {\it spin network} basis.  This is equivalent to
the choice of an orthonormal basis for the intertwining operator
\begin{equation}
    \sum_{\alpha_\mathbf{x}^k,\beta_\mathbf{x}^k=1}^{
    {\mathrm{dim}(j_\mathbf{x}^{k})}}
    \mathsf{I}_{\mathbf{x}}^{\Sg{\mathbf{\pi}_\mathbf{x}}}
    {}^\g{j_{\xm[1]}^{1},\ldots,j_{\xm[d]}^{d}}_{
    {\alpha}_{\xm[1]}^1,\ldots,{\alpha}_{\xm[d]}^d}
    {}_\g{j_\mathbf{x}^1,\ldots,j_\mathbf{x}^d}^{
    {\beta}_\mathbf{x}^1,\ldots,{\beta}_\mathbf{x}^d} ~\cdot~
    \mathsf{I}^\mathbf{x}_\Sg{\mathbf{\pi}_\mathbf{x}'}
    {}_\g{j_{\xm[1]}^{1},\ldots,j_{\xm[d]}^{d}}^{
    \alpha_{\xm[1]}^1,\ldots,\alpha_{\xm[d]}^d}
    {}^\g{j_\mathbf{x}^{1},\ldots,j_\mathbf{x}^{d}}_{
    \beta_\mathbf{x}^1,\ldots,\beta_\mathbf{x}^d} =
    \delta_\Sg{\mathbf{\pi}_\mathbf{x}'}^\Sg{\mathbf{\pi}_\mathbf{x}}
    ~~.
\label{eq4}
\end{equation}
As an example, for $SU(2)$ gauge group in $d=2$ dimensions the
spin-network basis elements can be characterized by three half integers
associated to each lattice site. On the other hand,
in $d=3$ dimensions six half integers are needed per lattice site (see Fig. 1).


\section{Matrix elements of the plaquette operator}

Using the integral (\ref{eq2}), it is straightforward to compute the 
action of the plaquette operator on the {\it spin-network basis}:
just take the trace of the intertwining operators. More specifically,
this consists of the evaluation of specific Wigner's $nJ$-symbols, which, in turn, involve
traces of $6J$-symbols. The final result is:
\begin{eqnarray} \label{eq5}
&& \langle \vec{\jmath}~',\mathbf{\vec{\pi}'}| U_{\mathbf{y},r,s} |
\vec{\jmath},\mathbf{\vec{\pi}}\rangle = \int \prod_{\mathbf{x},k}
dU_k(\mathbf{x})
~\overline{\psi_{\vec{\jmath}~',\mathbf{\vec{\pi}}'}(U)}
~\psi_{\vec{\jmath},\mathbf{\vec{\pi}}}(U) ~\cdot \\[2mm] &&
~~~~~~~~~~~ \cdot U^{-\!1}_s{}^{\tau_1}_{\tau_4}(\mathbf{y})
U^{-\!1}_r{}^{\tau_4}_{\tau_3}(\mathbf{y}+a \mathbf{e}_s)
U_s{}^{\tau_3}_{\tau_2}(\mathbf{y}+a \mathbf{e}_r)
U_r{}^{\tau_2}_{\tau_1}(\mathbf{y}) = \nonumber \\ && ~~~ = \int
\prod_{\mathbf{x}} \prod_{k=1}^{d} dU_k(\mathbf{x})
\sum_{\bar{\alpha}_\mathbf{x}^k,\bar{\beta}_\mathbf{x}^k=1}^{
{\mathrm{dim}(j_\mathbf{x}^k)}}
\sum_{\alpha_\mathbf{x}^k,\beta_\mathbf{x}^k=1}^{
{\mathrm{dim}(j_\mathbf{x}^{k\prime})}} ~\Rgen[j_\mathbf{x}^k
]{\bar{\beta}_\mathbf{x}^k}{
{\bar\alpha}_\mathbf{x}^k}(U_k^{-\!1}(\mathbf{x}))
~\Rgen[j_\mathbf{x}^{k\prime}]{{{\alpha}}_\mathbf{x}^k}{
{{\beta}}_\mathbf{x}^k}(U_k(\mathbf{x})) \cdot
\nonumber \\[2mm] && ~~~~~~~~~~~ \cdot
U^{-\!1}_s{}^{\tau_1}_{\tau_4}(\mathbf{y})
U^{-\!1}_r{}^{\tau_4}_{\tau_3}(\mathbf{y}+a \mathbf{e}_s)
U_s{}^{\tau_3}_{\tau_2}(\mathbf{y}+a \mathbf{e}_r)
U_r{}^{\tau_2}_{\tau_1}(\mathbf{y}) \cdot \nonumber \\[2mm] &&
~~~~~~~~~~~ \cdot
\mathsf{I}_{\mathbf{x}}^{\Sg{{\mathbf{\pi}_\mathbf{x}}}}
{}^\g{j_{\xm[1]}^1,\ldots,j_{\xm[d]}^d
}_{\bar{\alpha}_{\xm[1]}^1,\ldots,\bar{\alpha}_{\xm[d]}^d}
{}_\g{j_\mathbf{x}^1,\ldots,j_\mathbf{x}^d
}^{\bar{\beta}_\mathbf{x}^1,\ldots,\bar{\beta}_\mathbf{x}^d} \cdot
\mathsf{I}^\mathbf{x}_\Sg{\mathbf{\pi}_\mathbf{x}'}
{}_\g{j_{\xm[1]}^{1\prime},\ldots,j_{\xm[d]}^{d\prime}
}^{\alpha_{\xm[1]}^1,\ldots,\alpha_{\xm[d]}^d}
{}^\g{j_\mathbf{x}^{1\prime},\ldots,j_\mathbf{x}^{d\prime}
}_{\beta_\mathbf{x}^1,\ldots,\beta_\mathbf{x}^d} \nonumber
\end{eqnarray}
Because of the traces, the choice of an 
explicit basis is irrelevant, as expected. All the indices of the intertwiner
matrix elements are traced over their complex conjugate, except the
contractor in the lattice points $\mathbf{y}$, $\yp[r]$, $\yp[s]$ and
$\yp[r+s]$. The corresponding matrix elements are
\begin{equation} \label{eq6}
\langle \vec{\jmath}~',\mathbf{\vec{\pi}'}| U_{\mathbf{y},r,s} |
\vec{\jmath},\mathbf{\vec{\pi}}\rangle = ~\mbox{[Eq. (26) of ref.\ 1]} \;.
\end{equation}
Notice that once the intertwining matrices are specified, i.e., when
the Clebsh-Gordan coefficients are explicitly given, the matrix
elements are known.  In this way we have reduced the problem of the
computation of the matrix elements of the plaquette operator to the
computation of the trace of intertwining operators, i.e., of the trace
of generalized Clebsh-Gordan coefficients. Now, it is well known that
this is nothing more than the evaluation of specific Wigner's
$nJ$-symbols and this  can be
always reduced to the computation of a Wigner's $6J$-symbol.

In the case of SU(2) the Wigner's $6J$-symbols are completely known
and it is possible to find algebraic expressions for such matrix 
elements. In particular, in 2+1 dimensions and using the corresponding basis 
of Fig. 1 the matrix elements of the 
plaquette $U_{{\mathbf{y}},1,2}$ can be given. They are
different from zero only if all the six primed and un-primed
$j_{\mathbf{x}}^1$, $j_{\mathbf{x}}^2$, $\pi^1_{\xp[1]}$,
$j^2_{\xp[1]}$, $\pi^1_{\xp[2]}$, $j^1_{\xp[2]}$, differ by a half
integer for $\mathbf{x}\!=\!\mathbf{y}$.  Their explicit expression is
\begin{eqnarray} 
\nonumber \langle \mathbf{\vec{\jmath}~'},\mathbf{\vec{\pi}'}|
U_{{\mathbf{y}},1,2} | \mathbf{\vec{\jmath}},\mathbf{\vec{\pi}}\rangle
= \frac{ (-1)^{\sum_{i=1}^n \left(
\left|\epsilon_i-\epsilon_{i\!+\!1}\right| + \frac{ C^i_{\mathbf{y}}
}{2} \right)} }{ \sqrt{\prod_{i=1}^n \left( 2 X^i_{\mathbf{y}}+ 1
\right) \left( 2 Y^i_{\mathbf{y}}+ 1 \right) } }
\prod_{i=1}^n R\left[{ \begin{array}{cc} X^i_{\mathbf{y}} &
X^{i\!+\!1}_{\mathbf{y}} \\ Y^i_{\mathbf{y}} &
Y^{i\!+\!1}_{\mathbf{y}}
    \end{array},C^i_{\mathbf{y}} }\right] 
\end{eqnarray}
where $\epsilon_i={X}^i_{\mathbf{y}}-{Y}^i_{\mathbf{y}}=\pm
\frac{1}{2}$,
$$
\begin{array}{llllll}
{X}^1_{\mathbf{y}}\!=\!j_{\mathbf{x}}^1 ~, &
{X}^2_{\mathbf{y}}\!=\!j_{\mathbf{x}}^2 ~, &
{X}^3_{\mathbf{y}}\!=\!\pi^1_{\xp[2]} ~, &
{X}^4_{\mathbf{y}}\!=\!j^1_{\xp[2]} ~, &
{X}^5_{\mathbf{y}}\!=\!j^2_{\xp[1]} ~, &
{X}^6_{\mathbf{y}}\!=\!\pi^1_{\xp[1]} ~ \\
{Y}^1_{\mathbf{y}}\!=\!j_{\mathbf{x}}^{1\prime} ~, &
{Y}^2_{\mathbf{y}}\!=\!j_{\mathbf{x}}^{2\prime} ~, &
{Y}^3_{\mathbf{y}}\!=\!\pi^{1\prime}_{\xp[2]} ~, &
{Y}^4_{\mathbf{y}}\!=\!j^{1\prime}_{\xp[2]} ~, &
{Y}^5_{\mathbf{y}}\!=\!j^{2\prime}_{\xp[1]} ~, &
{Y}^6_{\mathbf{y}}\!=\!\pi^{1\prime}_{\xp[1]} \\
{C}^1_{\mathbf{y}}\!=\!\pi_{\mathbf{x}}^1 ~, &
{C}^2_{\mathbf{y}}\!=\!j_{\xm[1\!+\!2]}^1 ~, &
{C}^3_{\mathbf{y}}\!=\!j^2_{\xp[2]} ~, &
{C}^4_{\mathbf{y}}\!=\!\pi^1_{\xp[1\!+\!2]} ~, &
{C}^6_{\mathbf{y}}\!=\!j^1_{\xp[1]} ~, &
{C}^6_{\mathbf{y}}\!=\!j^2_{\xp[1\!-\!2]} ~,
\end{array}
$$ and $R\left[{ \begin{array}{cc} X^i_\mathbf{y} &
X^{i\!+\!1}_\mathbf{y} \\ Y^i_\mathbf{y} & Y^{i\!+\!1}_\mathbf{y}
    \end{array},C^i_\mathbf{y} }\right] 
$ is equal to
\[
\left\{ \begin{array}{lcl}
    \sqrt{\frac{1-2 C^i_\mathbf{y}
    +X^i_\mathbf{y}+X^{i\!+\!1}_\mathbf{y}
    +Y^i_\mathbf{y}+Y^{i\!+\!1}_\mathbf{y} }{2} ~ \frac{3+2
    C^i_\mathbf{y} +X^i_\mathbf{y}+X^{i\!+\!1}_\mathbf{y}
    +Y^i_\mathbf{y}+Y^{i\!+\!1}_\mathbf{y} }{2} } & &\text{if~}
    \left|\epsilon_i-\epsilon_{i\!+\!1}\right|=0 \\[3mm]
    \sqrt{\frac{1+2 C^i_\mathbf{y}
    +X^i_\mathbf{y}-X^{i\!+\!1}_\mathbf{y}
    +Y^i_\mathbf{y}-Y^{i\!+\!1}_\mathbf{y} }{2} ~ \frac{1+2
    C^i_\mathbf{y} -X^i_\mathbf{y}+X^{i\!+\!1}_\mathbf{y}
    -Y^i_\mathbf{y}+Y^{i\!+\!1}_\mathbf{y} }{2} } & &\text{if~}
    \left|\epsilon_i-\epsilon_{i\!+\!1}\right|=1
    \end{array}\right. 
\]

A straightforward application of this result is the computation of the
matrix elements of the LGT Hamiltonian operator
\begin{equation} \label{def:HAM}
\hat{H} = \frac{g^2}{2a^{d\!-\!2}} \sum_{{\mathbf{x}},k}
q_{\alpha\beta} E^\alpha_k({\mathbf{x}})E^\beta_k({\mathbf{x}}) +
\sum_P \frac{a^{d\!-\!4}}{g^2} \left[1 - \frac{U_P +
U_P^*}{2\mathrm{dim}(U)} \right],
\end{equation}
where $q_{\alpha\beta}$ is the Cartan metric. The sum over $P$ runs
over all unoriented plaquettes, $U_P$ being the plaquette variable.

In fact, the basis vectors (\ref{eq3}) are eigenstates of the kinetic
term, while the potential (magnetic) term is realized as a
multiplicative operator.  Namely
\begin{eqnarray} \label{eq:HspinNET}
&& \langle \vec{\jmath}~',{\mathbf{\vec{\pi}}'}| \hat{H} |
\vec{\jmath},{\mathbf{\vec{\pi}}}\rangle = \left( \frac{g^2}{2\
a^{d\!-\!2}} \sum_{\mathbf{x}}\sum_{k=1}^{d} C_2[j_{\mathbf{x}}^2] ~
+\frac{a^{d\!-\!4}}{g^2} N_P \right)
\delta_{\vec{\jmath}}^{\vec{\jmath}'}
\delta_{\mathbf{\vec{\pi}}}^{{\mathbf{\vec{\pi}}}'} + \\ &&~~~~ -
\frac{a^{d\!-\!4}}{2\ g^2\ {\rm dim}(U)} \sum_{\mathbf{y}}
\sum_{r<s=1..d} \bigg( \langle \vec{\jmath}~',{\mathbf{\vec{\pi}}'}|
U_
{{\mathbf{y}},r,s} | \vec{\jmath},{\mathbf{\vec{\pi}}}\rangle
  +\langle \vec{\jmath},{\mathbf{\vec{\pi}}}| 
   U_{{\mathbf{y}},r,s} | \vec{\jmath}~',{\mathbf{\vec{\pi}}'}\rangle
\bigg)
\nonumber 
\end{eqnarray}
where the only non diagonal terms are just the
afore computed expectation values of the plaquette operator.

\section*{Acknowledgments}
We want to thank E. Onofri and F. Di Renzo for helpful 
and enlightening discussions. Partial support from grants CONACyT 
No. 3141P-E9608 and DGAPA-UNAM IN100397 is also acknowledged.

\end{document}